\documentclass[twocolumn,aps,pra,showpacs,superscriptaddress,amsfonts,amssymb,amsmath]{revtex4-1}
\usepackage{CJK}
\usepackage{bm,amsmath,amssymb,amsfonts}
\usepackage{color}
\usepackage{dcolumn}
\usepackage{graphicx}
\usepackage{float}
\usepackage{hyperref}
\hypersetup{colorlinks=true,citecolor=blue,urlcolor=blue,linkcolor=blue}
\usepackage{ulem}
\usepackage{times}
\usepackage{subfigure}

\begin{document}

\title{Floquet engineering tunable periodic gauge fields and simulating real topological phases in cold alkaline-earth atom optical lattice}

\author{Wei Wang}
\affiliation{Department of Physics and Chongqing Key Laboratory for Strongly Coupled Physics, Chongqing University, Chongqing 401331, China}
\affiliation{Center of Modern Physics, Institute for Smart City of Chongqing University in Liyang, Liyang, 213300, China}

\author{Zheng Zhang}
\affiliation{National Laboratory of Solid State Microstructures and Department of Physics, Nanjing University, Nanjing 210093, China}
\author{Gui-Xin Tang}
\affiliation{Center of Ultra-precision Optoelectronic Instrument Engineering, Harbin Institute of Technology, Harbin 150001, China }
\author{Tao Wang}
\email{tauwaang@cqu.edu.cn}
\affiliation{Department of Physics and Chongqing Key Laboratory for Strongly Coupled Physics, Chongqing University, Chongqing 401331, China}
\affiliation{Center of Modern Physics, Institute for Smart City of Chongqing University in Liyang, Liyang, 213300, China}
\affiliation{State Key Laboratory of Quantum Optics and Quantum Optics Devices, Shanxi University, Taiyuan, 030006,  China}

\begin{abstract}
We propose to synthesize tunable periodic gauge fields via Floquet engineering cold alkaline-earth atoms in one-dimensional optical lattice.
The artificial magnetic flux is designed to emerge during the combined process of Floquet photon assisted tunneling and internal state transitions.
By varying initial phases of driving protocol, our proposal presents the ability to smoothly tune the periodic flux.
Moreover, we demonstrate that the effective two-leg flux ladder model can simulate one typical real topological insulator, which is described by the first Stiefel Whitney class and protected by the $PT$ symmetry.
Benefiting from the long lifetime of excited states of alkaline-earth atoms, our work opens new possibilities for exploiting the physics related to gauge fields, such as topological phases, in the current cold atom platform.

\end{abstract}

\maketitle

\section{INTRODUCTION}

Gauge fields, as well as the associated gauge theories, are crucial to modern physics.
In the standard model, complex gauge fields are necessary to mediate the interactions between elementary particles.
The application of strong magnetic fields to two-dimensional electronic systems has led to the discovery of topological matters\cite{Klitzing1980,Laughlin1981,Thouless1982,Laughlin1983,Tsui1982} that have been actively expanded over the last fifteen years\cite{Hasan2010,Qi2011,Chiu2016,Bansil2016,Armitage2018}.
Among those topological matters, Chern insulators have drawn tremendous attention for exploring the topological mechanisms beyond Landau levels and their potential application aspects,
which was first proposed by Haldane through introducing staggered fluxes threading the honeycomb lattice\cite{Haldane1988}.
Inspired by this spatial magnetic configuration, much of research is devoted to combining that with various lattice systems, whose interplay results in flat bands\cite{Ohgushi2000,Sun2011,Neupert2011,Chen2012,Ge2021,Guan2023,Lan2023}, anomalous quantum Hall effects\cite{Ohgushi2000,Xu2015,Zhang2017}, high Chern numbers\cite{Chen2012,Lan2023}, high-order Chern number\cite{Guan2023}, unique edge states\cite{He12022,He22022} and so on.
Extra effort is focused on investigating other periodic fluxes to explore similar phenomena such as high Chern numbers\cite{Liu2012}, redistribution of Chern numbers\cite{Wang2007,Sun2016}, chiral edge states\cite{Cai2019}.
Nevertheless, it is interesting to note that there is a kind of novel topological phase recently emerging from the periodic $\pi$ flux configuration\cite{Zhao2020,Shao2021,Chen2022,Chen2023,Zhang2023,Jiang2024,Zhao2021,Xue2022,Xue2023}.
Distinct from other precursors, their Brillouin zone, band topologies, edge states, symmetry groups and topological classifications are profoundly modified by the $Z_2$ projective algebra\cite{Zhao2020,Shao2021,Chen2022,Chen2023,Zhang2023,Jiang2024}.
It is naturally anticipated that more general periodic $U(1)$ gauge fields may extend the realm of intriguing topological phases\cite{Shao2021}.
However, how to engineer periodic gauge fields is still an open question to date.

Thanks to the unprecedentedly clean and controllable experimental system, cold atoms offer a unique platform for simulating and investigating the gauge fields\cite{Dalibard2011,Goldman2014R,Zhang22017,Aidelsburger2018,Weitenberg2021}.
One early and simple route for synthesizing effective magnetic fields involves rapidly rotating the cold gases, leveraging the analogy between the neutral atomic Coriolis force and the charged particles' Lorentz force\cite{Matthews1999,Madison2000,Abo-Shaeer2001,Engels2003,Schweikhard2004}.
Later, the scheme of laser-assisted tunneling is proposed to exploit Peierls phases which arise when the suppressed adjacent tunneling is resonantly restored by Raman transitions\cite{Ruostekoski2002,Jaksch2003,Yi2008,Gerbier2010,Celi2014,Aidelsburger2011}.
More elaborate dynamical driving technique--commonly referred to as Floquet engineering\cite{Eckardt2005,Lignier2007,Zenesini2009,Struck2011,Hauke2012,Jimenez-Garcia2012,Goldman2014,Bukov2016,Plekhanov2017,Struck2012,Eckardt2017,Weitenberg2021}--is employed to  attach gauge fields to varied lattice configurations such as triangular lattice, kagome lattice, hexagonal lattice\cite{Struck2011,Hauke2012,Plekhanov2017,Struck2013,Jotzu2014,Flaschner2016}.
The proven versatile tool can even engineer solenoid-type flux geometries\cite{Wang2018}.
On the other hand, cold alkaline-earth atoms(AEAs) offer unique advantages for simulating gauge fields by utilizing long-lived electronic excited states,
as the strongly suppressed spontaneous emission reduces related heating\cite{Gerbier2010,Mancini2015,Wall2016,Livi2016,Kolkowitz2017,Bromley2018}.
This useful low heating rate has allowed the application of Floquet engineering methods in optical lattices to advance recent experimental progress not only in quantum simulation but also in precision measurement\cite{Lu2021,Yin2022}.
These developments motivate us to explore the possibility of engineering periodic artificial gauge fields in such atomic platforms.

In this paper, we propose a feasible and efficient scheme for the generation of tunable periodic gauge fields by Floquet engineering of cold AEAs.
By designing an appropriate superlattice, we show that these atoms experience state-dependent potentials, and thus acquire net magnetic fluxes attaching the artificial "electronic" dimension
due to different Floquet photon assisted resonant precesses.
Intriguingly, it is found that the effective two-leg periodic flux model can exhibit the real topological phase.
This topological state is protected by $PT$ symmetry, the combined symmetry of spatial inversion $P$ and time reversal $T$, and characterized by the first Stiefel Whitney class.
Topological phase transition can occur when tuning the periodic flux.

The paper is organized as follows.
Section II serves as an introduction to our proposal and the corresponding time-dependent Floquet model.
In Sec.III, we discuss the effective Hamiltonian in case of the Floquet photon assisted resonant tunneling between adjacent sublattices, and transition between two levels of atoms.
According to that, the tunability of periodic artificial gauge potentials is then shown.
We discuss the topological properties of the effective model in Sec. IV.
Finally in Sec.V, we conclude with some discussions and remarks.

\section{proposal}

\begin{figure}[t]
\begin{center}
\includegraphics[height=3cm,width=8cm]{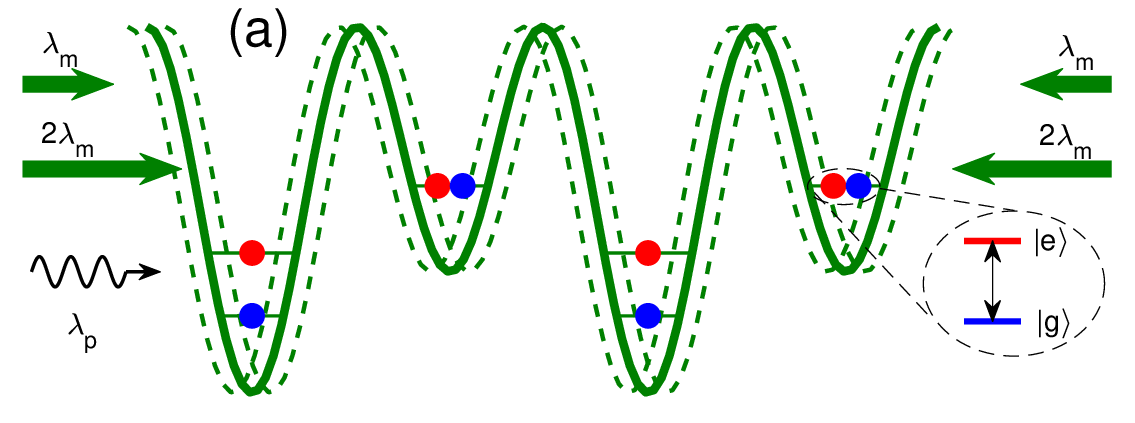}
\includegraphics[height=3cm,width=4cm]{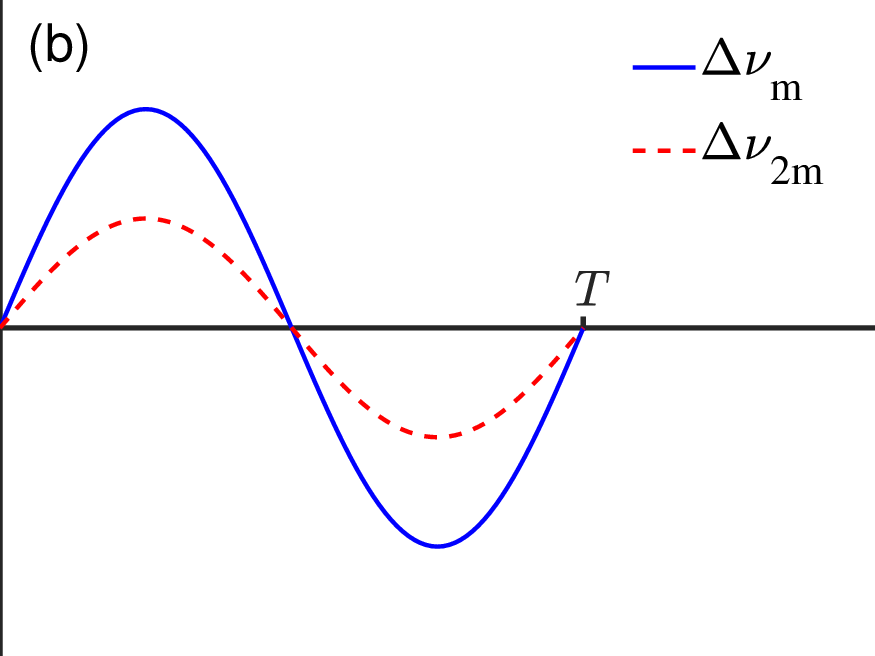}
\includegraphics[height=3cm,width=4cm]{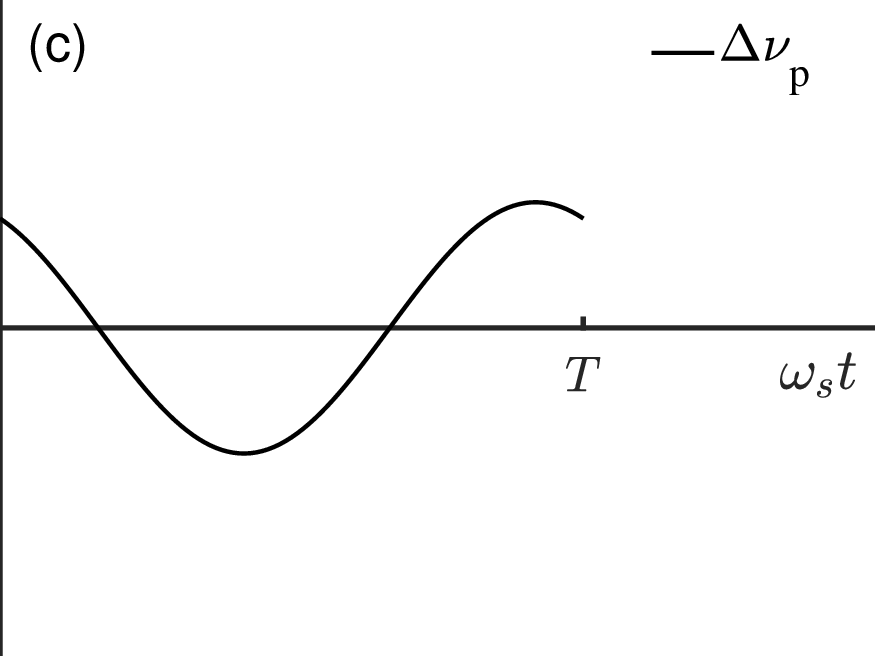}
\caption{Sketch of proposed setup. (a) Cold AEAs are trapped in shaking superlattice formed by two standing-wave laser fields with wavelength $\lambda_m$ and $2\lambda_m$.
Interacting with modulated pumping lasers, atoms with ground states $|g\rangle$ can be excited to higher energy levels $|e\rangle$.
(b) The periodic shaking of superlattice originates from the two coordinated frequency modulation function for the laser with magic wavelength $\Delta\nu_m=\nu_1\sin(\omega_s t)$ and the other $\Delta\nu_{2m}=\frac{\nu_1}2\sin(\omega_s t)$, denoted dy blue solid line and red dash line respectively.
(c) The modulation function to frequency of the pumping laser, $\Delta\nu_p(t)=\nu_2\cos(\omega_s t+\varphi)+\frac{\lambda_m}{2\lambda_p}\nu_1\sin(t)$, with setting $\nu_2=\nu_1$, $\varphi=\frac\pi3$ and $\frac{\lambda_m}{\lambda_p}=\frac{7\pi}{6}$.}\label{setup}
\end{center}
\end{figure}

Illuminated by the modulated pumping laser, we consider cold AEA gases confined in an one-dimensional driven superlattice, which is illustrated intuitively in Fig.\ref{setup}(a).
The pumping laser interrogates the transition between the ground states $|g\rangle$ and the excited states $|e\rangle$.
Such superlattice is formed by overlapping two 1D optical lattices with one at a magic wavelength $\lambda_m$ giving lattice depth $V_m$, and the other at the wavelength $2\lambda_m$ giving the depth $V_g$ and $V_e$ for the state $|g\rangle$ and $|e\rangle$ respectively.
To simplify our discussion, the superlattice and pumping laser are assumed to be driven simultaneously and the sine protocol is chosen, which can be achieved by acousto-optic modulators.

Define $\mathcal{P}_g$ and $\mathcal{P}_e$ as the projection operator to the ground states $|g\rangle$ and the excited states $|e\rangle$ respectively,
the driven state-dependent superlattice can be written as
\begin{eqnarray}\label{V}
V(x,t)&=&-\frac{V_m}{2}\cos(2k_m[x-X_L(t)])\nonumber\\
&&-\frac{V_g}{2}\mathcal{P}_g \cos(k_m[x-X_L(t)])\nonumber\\
&&-\frac{V_e}{2}\mathcal{P}_e \cos(k_m[x-X_L(t)]),
\end{eqnarray}
where the sinusoidal driven function is applied
\begin{eqnarray}\label{dL}
X_L(t)=\frac{\lambda_m}{2}\int^t_0 \nu_1\sin(\omega_s t) d\tau,
\end{eqnarray}
$\omega_s/2\pi$ denotes the driving frequency.
The right hand side of the first line in Eq.(\ref{V}) indicates that the atomic two levels feel the same lattice potential generated by the magic wavelength $\lambda_m$.
In contrast, the second and third lines describe different trapping conditions due to the laser with wavelength $2\lambda_m$.
Assuming identical spatial shaking $X_L(t)$ for the two lattice, the two driven protocols should be coordinated as shown in Fig.\ref{setup}(b).
Let $\nu_1$ be the frequency excursion of the modulation to the magic laser frequency.
Such condition can be achieved by just choosing the frequency excursion $\nu_1/2$ for the other one.

Considering another time-dependent modulation to the frequency of the pumping laser,
\begin{eqnarray}\label{dp}
\Delta\nu_p(t)=\nu_2\cos(\omega_s t+\varphi)+\frac{1}{\lambda_p}\dot{X}_L(t),
\end{eqnarray}
which is shown in Fig.\ref{setup}(c), the $|g\rangle\leftrightarrow|e\rangle$ transition can be described by the atom-laser couping matrix(under the rotating-wave approximation),
\begin{eqnarray}
W=\frac{\hbar}{2}\begin{pmatrix}
\delta & g e^{ik_p[x-X_p(t)]}\\
g e^{-ik_p[x-X_p(t)]}&-\delta
\end{pmatrix}\nonumber,
\end{eqnarray}
with
\begin{eqnarray}
X_p(t)=\lambda_p\int^t_0 d\tau\Delta\nu_p(\tau).
\end{eqnarray}
Here $g$ is the Rabi frequency, $\delta=\omega_0-\omega_p$ is the detunning,
$\omega_0$ is the frequency difference between $|g\rangle$ and $|e\rangle$, $\lambda_p$ and $k_p$ is the wavelength and wave number of pumping laser.
It is noteworthy that we have introduced an initial phase difference $\varphi$ between the two driven functions in Eq.(\ref{dL}), which is the key ingredient for engineering the periodic flux in our following discussions.
According to cold AEA experiments such as $^{173}$Yb\cite{Mancini2015,Livi2016} and $^{87}$Sr\cite{Kolkowitz2017,Lu2021,Yin2022}, neglecting atomic interaction in some suitable lattices is reasonable.
Under individual-particle approximation, then the external and internal motion of the atoms is governed by the Hamiltonian
\begin{eqnarray}\label{Ham}
H=\big[\frac{p^2}{2m}+V(x,t)\big]\otimes\hat{1}+W,
\end{eqnarray}
where $m$ and $p$ are the atomic mass and momentum, $\hat{1}$ is the identity operator associated with the internal atomic degrees of freedom.

\section{effective Hamiltonian and periodic gauge fields}
It is convenient to work in the frame of reference co-moving with the superlattice, into which we can transform the atomic motion by two steps of unitary transformation.
First, we define $U_1=\exp(\frac{i}{\hbar}X_L(t)p)$ and transform the Hamiltonian in Eq.(\ref{Ham}) by $H\to H_1=U_1 H U^\dag_1-i\hbar U_1\partial_t U^\dag_1$.
The shift of position in the potential $V(x,t)$ is compensated by $U_1 x U^\dag_1=x+X_L(t)$.
However, the extra term $-i\hbar U_1\partial_t U^\dag_1$ generates $-\dot{X_L}(t)p$, which means a shift of the momentum.
To cancel this extra term, we implement the other unitary transformation by $U_2=\exp(-\frac{i}{\hbar}m\dot{X}_L(t)x)$, which results in $p\to p+m\dot{X_L}(t)$.
Finally, the Hamiltonian $H_c= U_2 H_1 U^\dag_2-i\hbar U_2\partial_t U^\dag_2$ in the co-moving frame becomes
\begin{eqnarray}\label{Hamc}
&&H_c = \bigg[\frac{p^2}{2m}+V^\prime(x)+m\ddot{X}_L(t)x-\frac{m}{2}\dot{X_L}(t)^2\bigg]\otimes\hat{1} \nonumber\\
&&+\frac{\hbar}{2}\begin{pmatrix}
\delta & g e^{ik_p[x-X_p(t)+X_L(t)]}\\
g e^{-ik_p[x-X_p(t)+X_L(t)]} & -\delta
\end{pmatrix}
\end{eqnarray}
with the undriven superlattice potential
\begin{eqnarray}
V^\prime(x)&=&-\frac{V_1}{2}\cos(2k_L x)-\frac{V_g}{2}\mathcal{P}_g \cos(k_L x)\nonumber\\
&&-\frac{V_e}{2}\mathcal{P}_e \cos(k_L x).
\end{eqnarray}
Here the term $-\frac{m}{2}\dot{X_L}(t)^2$ can be ignored since it is a time-dependent energy shift.
In this laboratory frame, we can see now that the vibration of superlattice gives rise to two physical effects.
The first is the inertial force, given by $F(t)=-m\ddot{X}_L(t)$, which generates the energy term $m\ddot{X}_L(t)x$.
The second is the Doppler effect, related to the term $e^{ik_p X_L(t)}$.

In Wannier representation using the tight-binding approximation, the many-body Hamiltonian described by Eq.(\ref{Hamc}) can be formulated as
\begin{eqnarray}\label{Hamcc}
H_c&=&\sum_{l,\sigma=g,e} K_{\sigma} \big(b^\dag_{l,\sigma}a_{l,\sigma}+a^\dag_{l+1,\sigma}b_{l,\sigma}+h.c.\big) \nonumber\\
 &&+\frac{\hbar g}{2} e^{ik_p[X_L(t)-X_p(t)]}\sum_{l}( e^{i2l\theta} a^\dag_{l,e}a_{l,g}\nonumber\\
 &&+e^{i(2l+1)\theta} b^\dag_{l,e}b_{l,g}+h.c.)+H_F+H_p,
\end{eqnarray}
where $a_{l,\sigma}$($a^\dag_{l,\sigma}$) and $b_{l,\sigma}$($b^\dag_{l,\sigma}$) denote the fermionic annihilation(creation) operator for atoms occupying the Wannier state at the a and b sublattice of the $l$th site respectively,
$K_\sigma$ is the corresponding tunnelling amplitude, and $g$, $e$ label the internal states $|g\rangle$, $|e\rangle$ respectively.
Here the phase $\theta=\pi\lambda_m/\lambda_p$ that can be changed by adjusting the angle between the pumping laser and the superlattice\cite{Livi2016}.
The energies associated with the inertial force acting on atoms at different sites are given by
\begin{eqnarray}\label{HamF}
H_F=- F(t)\lambda_m \sum_{l,\sigma=g,e}\big[l a^\dag_{l,\sigma}a_{l,\sigma}+(l+\frac12) b^\dag_{l,\sigma}b_{l,\sigma}\big].
\end{eqnarray}
Let $\mu_{\alpha,\sigma}$ $(\alpha=a,b, \sigma=g,e)$ be the chemical potentials of atoms in superlattice.
Then the corresponding total potentials should include the detuning related energy, namely $\mu_{\alpha,\sigma}\pm \frac\hbar2\delta$.
Since we are interested in Floquet photon assisted resonant processes, we set these potentials equal to integer multiples of $\hbar\omega_s$,
\begin{eqnarray}\label{Hamp}
H_p&=&\hbar\omega_s \sum_{l}\big[ n_a a^\dag_{l,e}a_{l,e}+m_g b^\dag_{l,g}b_{l,g}\nonumber\\
   &&+(m_g+n_b)b^\dag_{l,e}b_{l,e}\big],
\end{eqnarray}
where the minimum is redefined as the zero point of potential energy.
This condition can be satisfied by choosing proper lattice laser power, pumping laser frequency and driving frequency.

We proceed to discuss the resonant situation.
In this case, we need to combine the potentials $H_p$ with the inertial force induced energies $H_F$ and the Doppler effect associated terms $e^{ik_p[X_L(t)-X_p(t)]}$ in Eq(\ref{Hamcc}).
Based on this consideration, we do a combined rotation transformation of $U_3=\exp (\frac{i}{\hbar}\int^t_0 H_F d\tau)$ and $U_4=\exp(\frac{i}{\hbar} H_pt)$,
which leads to the new Hamiltonian
\begin{eqnarray}\label{HamR}
H_R&=& U_4U_3 H_c U^\dag_3U^\dag_4-i\hbar U_4U_3\partial_t (U^\dag_3U^\dag_4) \nonumber\\
   &=&\sum_{l,\sigma=g,e} K_{\sigma} \big(\eta^\sigma_{b,a} b^\dag_{l,\sigma}a_{l,\sigma}+\eta^\sigma_{a,b} a^\dag_{l+1,\sigma}b_{l,\sigma}\big) \nonumber\\
    &&+\frac{\hbar g}{2} \sum_{l} \chi^l_{a} a^\dag_{l,e}a_{l,g}+\chi^l_{b}e^{i\theta} b^\dag_{l,e}b_{l,g}+h.c.,
\end{eqnarray}
where
\begin{eqnarray}\label{renor}
&&\eta^\sigma_{b,a} =\exp\big[-\frac{i}{2\hbar}\int^t_0 d\tau F(\tau)\lambda_m+im_\sigma\omega_st\big],\nonumber \\
&&\eta^\sigma_{a,b} =\exp\big[-\frac{i}{2\hbar}\int^t_0 d\tau F(\tau)\lambda_m-im_\sigma\omega_st\big],\nonumber \\
&&\chi^l_{\alpha}= \exp\big\{i2l\theta+ik_p[X_L(t)-X_p(t)]+in_\alpha\omega_st\big\},
\end{eqnarray}
with $\alpha$ denoting $a$ or $b$, and $m_e=m_g+n_b-n_a$.

\begin{figure}[t]
\begin{center}
\includegraphics[width=8cm]{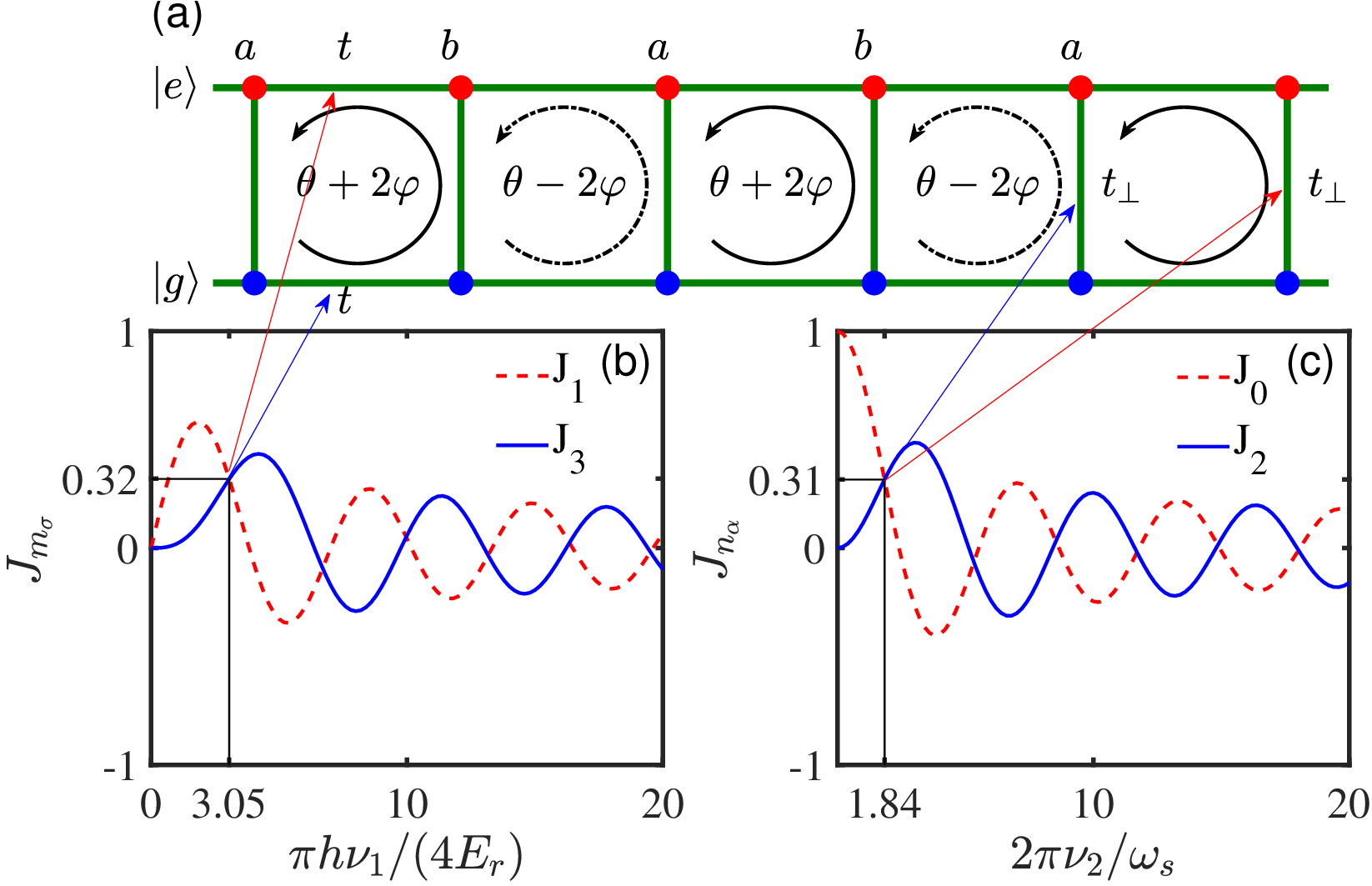}
\caption{Illustration of the effective two-leg ladder model with periodic gauge fields.
(a) A modulated pumping laser induces transition between ground states $|g\rangle$ and excited states $|e\rangle$, which can be viewed as two sites in the synthetic dimension.
The tunneling between sublattice is inhibited due to the potential offset, and can be restored with assistance of Floquet photons.
The Floquet photon assisted resonance pathways of one loop in each plaquette is accompanied by phases $\theta$ and $\varphi$, thus giving rise to periodic fluxes.
(b) The renormalized functions of the hopping amplitudes along the legs are Bessel functions, $J_1$ and $J_3$.
(c) The renormalized functions along the rungs, $J_0$ and $J_2$.}\label{effeHam}
\end{center}
\end{figure}

In a typical AEA experiment, for example $^{87}$Sr atoms\cite{Lu2021,Yin2022}, the driving frequency $\omega_s/2\pi$ can vary from several hundreds to several kilohertz, and the superlattice depth ranges from near zero to tens of recoil energies(the the recoil energy is defined as $E_r=h^2/(2m\lambda_m^2)$).
Therefore, we can adjust the experimental parameters to ensure that the driven frequency is larger than the inter-site tunneling amplitudes and the Rabi frequency, and then the high-frequency expansion can be applied safely.
In our proposal, the periodical modulation is sin signal function, as shown in Eq.(\ref{dL}) and Eq.(\ref{dp}).
We make use of the Bessel function expansion, $e^{iz\sin\beta}=\sum_k J_k(z) e^{ik\beta}$, to replace the driven modulation terms in Eq.(\ref{renor}).
Keeping only the resonant processes and neglecting other rapidly oscillating terms, the renormalized parameters of Eq.(\ref{renor}) can be approximately expressed as
\begin{eqnarray}\label{renormal}
&&\eta^\sigma_{b,a}\approx J_{-m_\sigma}(\Gamma_1),\nonumber \\
&&\eta^\sigma_{a,b}\approx J_{m_\sigma}(\Gamma_1),\nonumber \\
&&\chi^l_{\alpha}\approx J_{n_\alpha}(\Gamma_2)e^{2il\theta-in_\alpha\varphi},
\end{eqnarray}
where $\Gamma_{1}=\pi h\nu_{1}/(4E_r)$ and $\Gamma_{2}=2\pi\nu_{2}/\omega_s$.
Finally, substituting Eq.(\ref{renormal}) into Eq.(\ref{HamR}), we obtain a time-independent effective Hamiltonian.
By employing internal atomic degrees of freedom as the extra lattice dimension, it can be regarded as a two-leg flux ladder model depicted in Fig.\ref{effeHam}(a).

From the last line in Eq.(\ref{renormal}), we can see that phases $n_\alpha\varphi$, and thus the relevant gauge fields, emerge from the internal state resonant transition with the assistance of Floquet photons.
As for this artificial gauge field, the physical gauge-invariant quantity is the phase accumulated on an elementary loop per plaquette.
Taking into account the phase $\theta$, the total magnetic flux through each plaquette is $\theta\pm(n_a-n_b)\varphi$.
One should note that $J_{-k}=-J_{k}$ for odd $k$, which implies an additional $\pi$ phase.
To simplify the analysis, we always choose $m_{\sigma}$ and $n_\alpha$ such that the total extra phases can be eliminated.
Take a particular case of $n_a=2$, $n_b=0$, $m_g=3$, $m_e=1$ as an example, the accumulated phase is $\theta\pm2\varphi$.
Now it is clear that our scheme can simulate the periodic $U(1)$ gauge fields with two plaquette as one period.
Recall that $\theta$ is induced by incommensurate ratio between wavelengths of the lattice and the pumping lasers, it can be tuned by adjusting the angle between those lasers.
Moreover, $\varphi$ is the initial phase of the sinusoidal driving function and is fully controllable\cite{Lu2021}.
Thus our scheme offers a feasible method to tune the periodic gauge potentials.

Finally, we discuss the modulation of hopping amplitudes in the effective model.
The Bessel functions from Eq.(\ref{renormal}) renormalize the nearest-neighbor hopping amplitudes along legs as $K_{\sigma}J_{m_\sigma}(0.25\pi h\nu_{1}/E_r)$ and along rungs as $\frac{1}{2}\hbar g J_{n_\alpha}(2\pi\nu_{2}/\omega_s)$, respectively.
Independent of the phases $\theta$, $\varphi$, such simple parameter-dependent function forms result in their individual controllability via varying the driving amplitudes $\nu_1$, $\nu_2$.
While the demonstrated controllability offers exciting possibilities, it is essential to first consider the widely studied cases with equal hopping amplitudes along legs and along rungs.
Fig.\ref{effeHam}(b) shows the Bessel functions $J_{m_\sigma}$ for $m_g=3$ and $m_e=1$.
The crossing points $J_{3}(\Gamma_{1})=J_{1}(\Gamma_{1})$ indicate the equal tunneling amplitudes along legs if considering $K_{g}\approx K_{e}$.
Similarly, Fig.\ref{effeHam}(c) illustrates the tuned hopping amplitudes across rungs, with equal points clearly visible.

\section{Topological phase}

Now we investigate topological phase of the ladder model caused by the gauge fields.
For clarity, we rewrite the effective Hamiltonian for the case of $m_s=3$, $m_p=1$, $n_a=2$, $n_b=0$,
\begin{eqnarray}\label{RHam}
H_{eff}&=& t \sum_{l} \bigg[ e^{i(\theta+\phi)} b^\dag_{l,s}a_{l,s}+ e^{i(\theta-\phi)}a^\dag_{l+1,s}b_{l,s}\nonumber\\
       &&+b^\dag_{l,p}a_{l,p}+ a^\dag_{l+1,p}b_{l,p} \bigg ]\nonumber\\
    &&+t_{\bot}\sum_{l} \bigg[ a^\dag_{l,p}a_{l,s}+ b^\dag_{l,p}b_{l,s}\bigg]+h.c.
\end{eqnarray}
where $t=0.32K_{\sigma}$, $t_{\bot}=0.16\hbar g$, $\phi=2\varphi$, and the proper gauge transformation is preformed.
When $\phi=0$, namely the uniform flux scenario, previous research has demonstrated that no topological states are present in this 1D system\cite{Hugel2014}.
When $\theta+\phi=0$ and $\theta-\phi=\pi$, recent work by Zhao's group shows that nontrivial topological states manifest, and the topological invariant can be entirely determined by
the projective symmetry algebra\cite{Jiang2024}.
For $\phi/2\pi$ being the rational number, Sun\cite{Sun2016} has studied the particular case of periodic fluxes with period three, $\Omega_1=\theta+2\pi/3$, $\Omega_2=\theta+4\pi/3$, $\Omega_3=\theta+2\pi$, where the 1D topological invariant was elucidated with the help of the Chern numbers of the corresponding extended 2D system.

To explore the effects of general periodic gauge fields in Eq.(\ref{RHam}), we notice that this model exhibits combined symmetry of inversion $P$ and time-reversal $T$:
\begin{eqnarray}\label{PT}
PTH_{eff}(PT)^{-1}=H_{eff}.
\end{eqnarray}
In momentum space, the Hamiltonian and $PT$ operator are respectively represented by
\begin{eqnarray}\label{Hamk}
H_{eff}(k)&=&t\begin{pmatrix} 1+\cos(k) & 0\\ 0 & \cos(\theta+\phi)+\cos(k-\theta+\phi) \end{pmatrix} \otimes\sigma_1\nonumber\\
&&+t\begin{pmatrix}\sin(k) & 0\\ 0 & \sin(\theta+\phi)+\sin(k-\theta+\phi) \end{pmatrix} \otimes\sigma_2\nonumber\\
&&+t_{\bot} \tau_1\otimes\sigma_0, \nonumber\\
PT&=&\tau_0 \otimes\sigma_1 \mathcal{K},
\end{eqnarray}
where $\tau_i$, $\sigma_i$ denote Pauli matrices and $\mathcal{K}$ the complex conjugation.
Since $(PT)^2=1$, the $PT$ operator can be transformed into $\mathcal{K}$ by a unitary transformation, and the corresponding Hamiltonian $H_{eff}(k)$ will be required to be real in this basis.
When the number of ouccpied bands is one or three, the ground state is classified by the first Stiefel-Whitney class and characterized by a $\mathbb{Z}_2$-valued topological invariant
\cite{Ahn2018,Fang2015}. The topological invariant $\nu$ can be formulated with the help of the Wilson loop.
Introducing the nonabelian gauge connection
$[\mathcal{A}(k)]_{ab}=\langle u_a(k)|\partial_k|u_b(k)\rangle$, where $|u_a(k)\rangle$, $|u_b(k)\rangle$ are the wave functions of occupied bands, then
the Wilson loop is constructed as
\begin{eqnarray}\label{W}
W=P\exp(\int^{\pi}_{-\pi} \mathcal{A}(k) dk),
\end{eqnarray}
where $P$ indicate path ordering. $\nu$ is defined by the determinate of $W$,
\begin{eqnarray}\label{nu}
(-1)^{\nu}&=&\det W=\Pi_j e^{i\alpha_j}, \nonumber\\
\nu&=&\frac1\pi\sum_j\alpha_j \quad \mod  2,
\end{eqnarray}
where $e^{i\alpha_j}$ are eigenvalues of $W$. 
Due to the $PT$ symmetry, $\nu$ can only take value 0 or $1$ \cite{Ahn2018}.
\begin{figure}[t]
\begin{center}
\includegraphics[width=8cm]{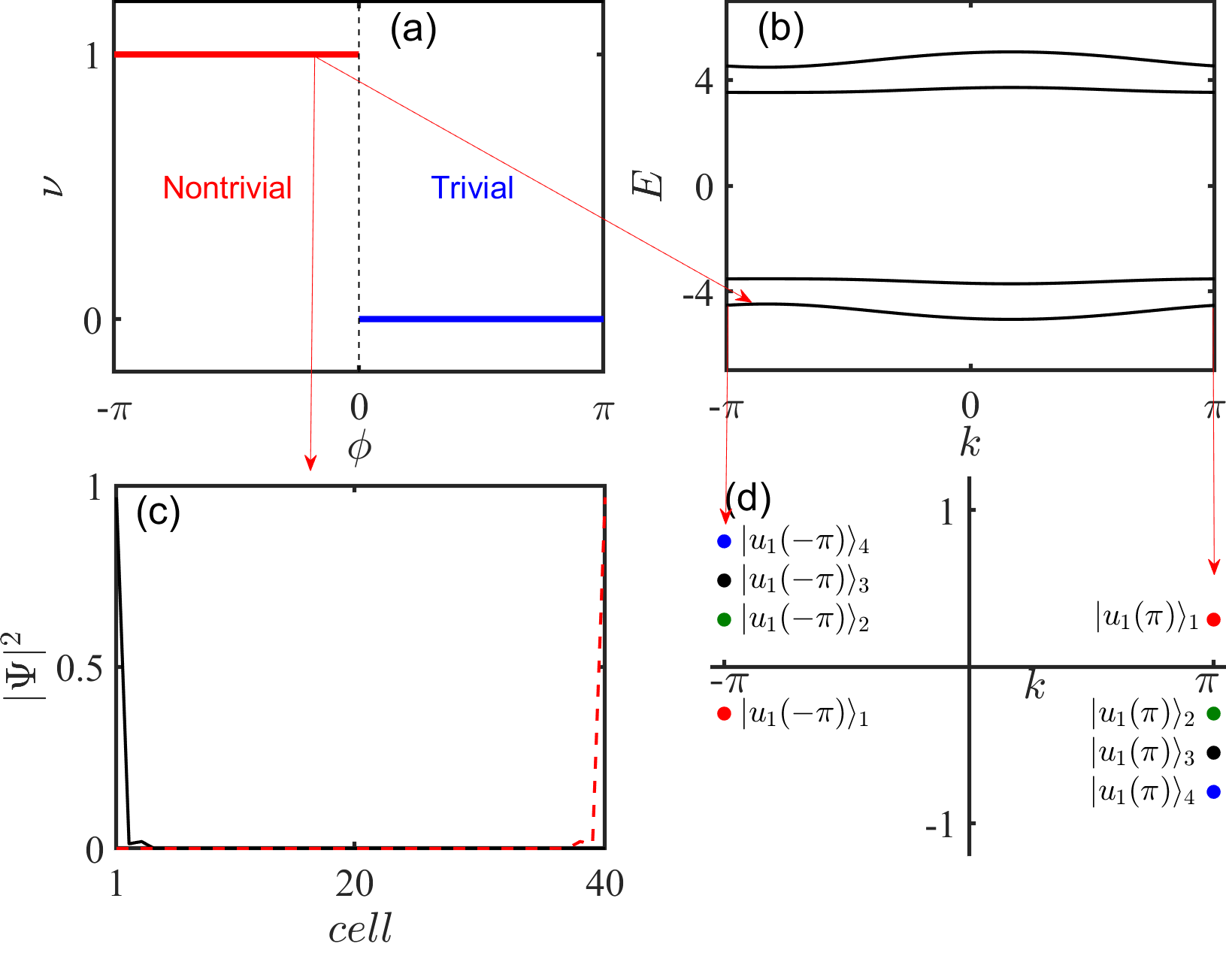}
\caption{(a) Topological phase diagram and $Z_2$ invariant $\nu$ versus $\phi$, where $\theta=7\pi/6$ and $t_{\bot}/t=4$.
(b) Dispersion relation with the same parameter of (a) but $\phi=-\pi/4$, where the topological number of the lowest band $\nu=1$.
(c) A pair of edge states corresponding to $\nu=1$ in (a) with $\phi=-\pi/4$, which is calculated for 40 unit cells in open boundary conditions.
(d) An intuitive illustration of the non-orientable state in the lowest band on the Brillouin zone boundaries $k=\pm\pi$.}\label{topo}
\end{center}
\end{figure}

Taking $^{87}$Sr atoms as an example, $\theta$ can be approximated as $7\pi/6$ if the pumping laser is parallel to the lattice laser\cite{Kolkowitz2017,Yin2022}.
We set $t_{\bot}/t$ as 4, and restrict $\phi$ within the interval $[-\pi,\pi]$ because of its periodicity.
Applying $Z_2$ invariant formulas in Eq.(\ref{W}) and Eq.(\ref{nu}) to the lowest filled band of the Hamiltonian in Eq.(\ref{Hamk}), one finds that $\nu=1$ for $-\pi<\phi<0$ and is trivial otherwise, as shown in Fig.\ref{topo}(a).
The topological number can be explained as follows\cite{Ahn2018}.
Considering the lowest band of energy spectrum[see Fig.\ref{topo}(b)], if we impose real conditions on the bulk wave function over the Brillouin zone,
the wave function can be made smooth over $-\pi<k<\pi$, and glued on the boundaries $k=\pm\pi$ but with an orientation-reversing transition function.
The transition function equaling to one indicates that the state is orientable and $\nu=0$, while minus one indicates that the state is non-orientable and $\nu=1$, 
as intuitively depicted in Fig.\ref{topo}(d).
Fig.\ref{topo}(c) shows the probability of a pair of edge states located inside the energy gap separating the second band form the lowest one, calculated for 40 unit cells with  open boundary conditions when $\phi=-\pi/4$.
When $\phi=0$, the system simplifies to the case of uniform fluxes, and the number of sites per unit cell reduces from four to two. 
This reduction indicates that the energy gap should close when $\phi=0$, which corresponds to a topological phase transition.

Additionally, we note that a similar discussion of topological properties can be conducted for three occupied bands, yielding the same results as that for one occupied bands.

\section{Conclusions}

In conclusion, we present a simple and feasible proposal for engineering a two-leg ladder model with periodic gauge fields based on driven cold AEA optical lattice systems.
The periodic gauge field is widely controllable by independently changing the parameters of the driven protocol. Our proposal can simulate real topological phase described by the
first Stiefel-Whitney class, and demonstrate topological phase transition induced by gauge fields.
Our scheme utilizes the long-lived electronic internal state, and thus offers a highly promising candidate for future experimental implementation and observation.
We hope these features of this work could enrich the gauge field-related topological research both in theory and in experiments.

Recently, the time-dependent synthetic gauge potentials has been theoretically confirmed as the critical ingredient for realizing tailored dynamical evolution of quasiparticles, quasiholes\cite{Wang2018,Raciunas2018,Wang2022}, wave packets\cite{Yilmaz2018,Lelas2021}, as well as adiabatic state preparation\cite{Wang2021}.
Another application of our proposal is expected for those dynamical process studies.
\section*{Acknowledgements}
This work was supported by the National Science Foundation of China under Grant No. 12274045. T. Wang acknowledges funding
from the National Science Foundation of China under Grants No.12347101 and funding from the Program of State Key Laboratory of Quantum Optics and Qauntum Optics Devices (Grant No. KF202211).


\end{document}